\documentclass[aps,prl,twocolumn,showpacs,groupedaddress]{revtex4-1}
\usepackage{amssymb}
\usepackage{mathrsfs}
\usepackage{amsmath}
\usepackage{mathtools}
\usepackage{pstricks}
\usepackage{color}
\usepackage{graphicx}
\usepackage{array}
\usepackage{tabularx}
\usepackage{physics}
\usepackage{float}

\begin{document}


\title{\bf Experimental validation of nonextensive statistical field theory: applications to manganites}



\author{P. R. S. Carvalho}
\email{prscarvalho@ufpi.edu.br}
\affiliation{\it Departamento de F\'\i sica, Universidade Federal do Piau\'\i, 64049-550, Teresina, PI, Brazil}




\begin{abstract}
In this Letter we validate experimentally the nonextensive statistical field theory, a new general field-theoretic approach introduced recently in the literature. With such an approach, we are capable of computing the critical properties of nonextensive systems undergoing continuous phase transitions, belonging to the new generalized O($N$)$_{q}$ universality class. We compare the nonextensive critical indices values evaluated from the theory with those already obtained in literature through experiments for various distinct manganites presenting nonconventional critical behavior. The agreement is satisfactory, whose relative errors are $< 5\%$ for the most of manganites employed and as better as $|1 - q| < 1$ which is the limit of validity of the theory. We present the physical interpretation of the experimental results and of the theory through the general physical interpretation of the $q$-parameter.
\end{abstract}


\maketitle


\par \textit{Introduction}.---Physics is one of the most successful branches of science. In fact, such a success was attained in describing many physical phenomena both qualitatively and quantitatively with high precision along the last decades. That success was achieved on the basis of the main philosophical idea called reductionism \cite{AnalysisFullness}. This idea asserts that the behavior of a physical system as a whole can be obtained from the knowledge of the behavior of its parts. Moreover, in the last years much attention has turned to develop strategies to comprehend physical systems called complex systems \cite{P.Anderson}, for which reductionism is not capable of explaining the corresponding experimental results. Its behavior can be described by considering not just its parts but their interactions as well. In general, complex systems are characterized by a large number of interacting constituents, nonlinearities, inhomogeneities, competition etc.. For studying such systems in the case of ones undergoing continuous phase transitions, some new general field-theoretic approach for computing their critical properties was introduced recently in the literature, namely nonextensive statistical field theory (NSFT) \cite{CARVALHO2022137147}. With that approach, which generalizes the extensive Boltzmann-Gibbs one \cite{Gibbs}, valid for $|1 - q| < 1$, they were computed both static and dynamic nonextensive critical exponents for systems, belonging to the new generalized O($N$)$_{q}$ universality class, like nonextensive Ising, XY, Kosterlitz-Thouless, Heisenberg, self-avoiding random walk, spherical \cite{CARVALHO2022137147} (for $\phi^{4}$ theory) and percolation, Lee-Yang edge singularity \cite{CARVALHO2022137284} (for $\phi^{3}$ theory). The nonextensive $q$-parameter was interpreted as one encoding effective interactions among the many degrees of freedom of a given system \cite{CARVALHO2022137147}. Now, besides depending on the dimension $d$, $N$ and symmetry of some $N$-component order parameter and if the interactions present are of short- or long-range type, the nonextensive critical exponents also depend on $q$ and satisfy to the same scaling relations as those satisfied by extensive critical indices as shown in Ref. \cite{CARVALHO2022137147}. We recover the extensive results obtained by Kenneth Wilson \cite{PhysRevLett.28.240} in the limit $q\rightarrow 1$. Kenneth Wilson's field-theoretic approach has been applied for determining precise values of extensive critical exponents \cite{PhysRevLett.28.240}. In fact, before that development we were constrained to employ only inaccurate approximations as the Landau one \cite{Callen}. In that approximation the values of the critical exponents are determined by considering only large length scales effects, which is approached by thermodynamics. Recently, it was shown that when we try to generalize thermodynamics to the nonextensive realm, the resulting theory can be mapped into its extensive counterpart by a transformation of variables \cite{PhysRevLett.88.020601}. As thermodynamics deals with physical phenomena only at large length scales, the effect of nonextensivity can not appear at such scales. So we expect that nonextensivity is associated to physical phenomena occurring just at small length scales, which is the scenario of statistical mechanics and is expressed as radiative loop corrections as shown recently in Ref. \cite{CARVALHO2022137147}. The predictions of NSFT have shown excellent agreement with those obtained from computer simulations for the static and dynamic critical exponents for the nonexetnsive two-dimensional Ising model \cite{PhysRevE.102.012116}. The purpose of this Letter is to show that the predictions of NSFT also agree with those obtained from experimental measurements on nonconventional real systems like manganites, \emph{e. g.}  La$_{0.67}$Sr$_{0.33}$MnO$_{3}$\cite{MNEFGUI2014193}. NSFT thus shows to be a generalization of Kenneth Wilson's extensive field-theoretic renormalization group in momentum space \cite{PhysRevLett.28.240} to describe nonconventional real or nonextensive systems.  

\par \textit{Nonextensive critical exponents}.---The NSFT is some perturbative Euclidean quantum field theory, where the generating functional is given by
\begin{eqnarray}\label{huyhtrjisd}
&& Z[J] = \mathcal{N}^{-1}\exp_{q}\left[-\int d^{d}x\mathcal{L}_{int}\left(\frac{\delta}{\delta J(x)}\right)\right]\times \nonumber \\&& \int\exp\left[\frac{1}{2}\int d^{d}xd^{d}x^{\prime}J(x)G_{0}(x-x^{\prime})J(x^{\prime})\right]
\end{eqnarray}
where $\exp_{q}(-x) = [1 - (1 - q)x]_{+}^{1/(1 - q)}$ is the $q$-exponential function \cite{Tsallis1988}($[y]_{+} \equiv y\theta(y)$ and $\theta(y)$ is the Heaviside step function). The constant $\mathcal{N}$ is determined from $Z[J=0] = 1$. In Refs. \cite{CARVALHO2022137147,CARVALHO2022137284} we have employed the Escort dictribution \cite{Astrophys.Space.Sci.} (Eq. (\ref{huyhtrjisd}) raised to a power of $q$). Once it has been show that this distribution is inconsistent \cite{Abe_2012}, in this work we have to employ the Eq. (\ref{huyhtrjisd}). Then, we can compute the static nonextensive critical exponents for O($N$) symmetry for $\phi^{4}$ theory through six distinct and independent methods in dimensions $d = 4 - \epsilon$ \cite{CARVALHO2022137147}. Thus we obtain two independent nonextensive critical indices valid for all loop orders, namely $\eta_{q}$ and $\nu_{q}$. They are given by 
\begin{eqnarray}\label{etaphi4}
\eta_{q} = \eta + \frac{1 - q}{q}\frac{(N + 2)\epsilon^{2}}{2(N + 8)^{2}},  
\end{eqnarray}
\begin{eqnarray}\label{nuphi4}
\nu_{q} = \nu + \frac{1 - q}{q}\frac{(N + 2)\epsilon}{4(N + 8)},
\end{eqnarray}
where $\eta$ and $\nu$ are the corresponding extensive critical exponents valid for all loop levels. The dynamic nonextensive critical exponent assumes the form 
\begin{eqnarray}\label{z}
z_{q} = z + \frac{1 - q}{q}\frac{[6\ln(4/3) - 1](N + 2)}{2(N + 8)^{2}}  \epsilon^{2}.
\end{eqnarray}

\par \textit{Comparison with experimental measurements}.--- Now we compare the critical exponents obtained from experimental measurements for various manganites with those predicted by NSFT of this work in Tables \ref{tableexponentsE3N3}-\ref{tableexponentsE3N1}. We have employed the extensive values $\eta = 0.061(8)$, $\nu = 0.689(2)$, $\beta = 0.365(3)$, $\gamma = 1.336(4)$ \cite{PHAN2010238} and $\eta = 0.030(4)$, $\nu = 0.630(1)$, $\beta = 0.325(2)$, $\gamma = 1.241(2)$ \cite{PHAN2010238} for Heinsenberg and Ising extensive systems, respectively, obtained from Eqs. (\ref{etaphi4})-(\ref{nuphi4}) and scaling relations among them \cite{CARVALHO2022137147}. 

\par \textit{Physical interpretation of the results}.---In Tables \ref{tableexponentsE3N3}-\ref{tableexponentsE3N1}, the values of the referred critical exponents differ from those for extensive Heinsenberg and Ising universality classeses \cite{PHAN2010238}, respectively, withing the margin of error. So manganites present nonconventional behavior \cite{Magnetochemistry.Turki,KHELIFI2014149,PhysRevB.75.024419,OMRI20123122,Ghosh_2005,doi:10.1063/1.2795796,GHODHBANE2013558,J.Appl.Phys.A.Berger,PhysRevB.68.144408,BenJemaa,PhysRevB.70.104417,PhysRevB.79.214426,J.Appl.Phys.Vasiliu-Doloca,YU2018393,PhysRevB.92.024409,ZHANG2013146,PHAN201440,RSCAdvJeddi,HCINI20152042,Phys.SolidStateBaazaoui}. By considering NSFT predictions of this work, they belong to the new generalized $q$-Heinsenberg ($N = 3$) and $q$-Ising ($N = 1$) universality classes, respectively, and the agreement with the experimental results is satisfactory, within some margin of error $< 5\%$ for the most of manganites employed and as better as $|1 - q| < 1$ which is the limit of validity of the theory. Within the margin of error, in the interval $1 < q < 2$ ($0 < q < 1$), the indices values decrease (increase), when compared to their corresponding extensive values (for $q = 1$). Also, within the margin of error, the critical temperatures decrease (increase) as the critical indices values increase (decrease) as can be seen in Table \ref{tablecriticalTC}, where we employ a few manganites with different concentrations.

\begin{table}[H]
\caption{Results for both static nonextensive critical exponents to $3$d ($\epsilon = 1$) nonextensive materials ($q \neq 1$) Heisenberg ($N = 3$) systems, obtained from experiment through Modified Arrott (MA) plots \cite{PhysRevLett.19.786}, Kouvel-Fisher (KF) method \cite{PhysRev.136.A1626} and NSFT of this work.}
\begin{tabular}{ p{5.6cm}p{1.3cm}p{1.3cm}  }
 \hline
 \hline
  & $\beta_{q}$ & $\gamma_{q}$    \\
 \hline
 La$_{0.67}$Sr$_{0.33}$MnO$_{3}$\cite{MNEFGUI2014193}MAP  &   0.333(8) &  1.325(1)    \\
 $q$ = 1.9 This work  &  0.334(3) &  1.238(6)     \\
 \hline
 Pr$_{0.77}$Pb$_{0.23}$MnO$_{3}$\cite{PhysRevB.75.024419}MAP  &   0.343(5) &  1.357(20)     \\
 Pr$_{0.77}$Pb$_{0.23}$MnO$_{3}$\cite{PhysRevB.75.024419}KF  &   0.344(1) &  1.352(6)     \\
 $q$ = 1.5 This work  &  0.343(3) &  1.267(7)     \\
 \hline
 AMnO$_{3}$\cite{OMRI20123122}MAP  &   0.355(7) &  1.326(2)     \\
 AMnO$_{3}$\cite{OMRI20123122}KF  &   0.344(5) &  1.335(2)     \\
 $q$ = 1.2 This work  &  0.354(3) &  1.301(7)     \\
 \hline
 Nd$_{0.7}$Pb$_{0.3}$MnO$_{3}$\cite{Ghosh_2005}MAP  &   0.361(13) &  1.325(1)           \\
 Nd$_{0.7}$Pb$_{0.3}$MnO$_{3}$\cite{Ghosh_2005}KF  &   0.361(5) &  1.314(1)           \\
 $q$ = 1.1 This work  &  0.360(3) &  1.318(7)     \\
 \hline
 LaTi$_{0.2}$Mn$_{0.8}$O$_{3}$\cite{doi:10.1063/1.2795796}KF  &   0.359(4) &  1.280(10)     \\
 $q$ = 1.1 This work  &  0.360(3) &  1.318(7)     \\
 \hline
 La$_{0.67}$Sr$_{0.33}$Mn$_{0.95}$V$_{0.05}$O$_{3}$\cite{MNEFGUI2014193}MAP  &   0.358(5) &  1.381(4)  \\
 $q$ = 1.1 This work  &  0.360(3) &  1.318(7)     \\
 \hline
 \hline
 $q$ = 1.0 \cite{GHODHBANE2013558} &  0.365(3) &  1.336(4)     \\
 \hline
 \hline
 La$_{0.67}$Sr$_{0.33}$Mn$_{0.95}$V$_{0.10}$O$_{3}$\cite{MNEFGUI2014193}MAP  &   0.367(3) &  1.414(2)  \\
 $q$ = 0.95 This work  &  0.369(3) &  1.347(7)     \\
 \hline
 La$_{0.67}$Ca$_{0.33}$MnO$_{3}$\cite{J.Appl.Phys.A.Berger}  &  0.368(3) &  1.384(17)  \\
 $q$ = 0.95 This work  &  0.369(3) &  1.347(7)     \\
 \hline
 Nd$_{0.85}$Pb$_{0.15}$MnO$_{3}$\cite{Ghosh_2005}MAP  &  0.372(1) &  1.340(30)  \\
 Nd$_{0.85}$Pb$_{0.15}$MnO$_{3}$\cite{Ghosh_2005}KF  &  0.372(4) &  1.347(1)  \\
 $q$ = 0.9 This work  &  0.373(3) &  1.360(7)     \\
 \hline
 Nd$_{0.6}$Pb$_{0.4}$MnO$_{3}$\cite{PhysRevB.68.144408}KF  &   0.374(6) &  1.329(3)  \\
 $q$ = 0.9 This work  &  0.373(3) &  1.360(7)     \\
 \hline
 La$_{0.67}$Sr$_{0.33}$Mn$_{0.95}$V$_{0.15}$O$_{3}$\cite{MNEFGUI2014193}MAP  &   0.375(3) &  1.355(6)  \\
 $q$ = 0.9 This work  &  0.373(3) &  1.360(7)     \\
 \hline
 La$_{0.67}$Ba$_{0.22}$Sr$_{0.11}$MnO$_{3}$\cite{BenJemaa}MAP  &   0.378(3) &  1.388(1)   \\
 La$_{0.67}$Ba$_{0.22}$Sr$_{0.11}$MnO$_{3}$\cite{BenJemaa}KF  &   0.386(6) &  1.393(4)   \\
 $q$ = 0.85 This work  &  0.378(3) &  1.372(7)     \\
 \hline
 LaTi$_{0.95}$Mn$_{0.05}$O$_{3}$\cite{doi:10.1063/1.2795796}KF  &   0.378(7) &  1.290(20)     \\
 $q$ = 0.85 This work  &  0.378(3) &  1.372(7)     \\
 \hline
 LaTi$_{0.9}$Mn$_{0.1}$O$_{3}$\cite{doi:10.1063/1.2795796}KF  &   0.375(5) &  1.250(20)     \\
 $q$ = 0.85 This work  &  0.378(3) &  1.372(7)     \\
 \hline
 LaTi$_{0.85}$Mn$_{0.15}$O$_{3}$\cite{doi:10.1063/1.2795796}KF  &   0.376(3) &  1.240(10)     \\
 $q$ = 0.85 This work  &  0.378(3) &  1.372(7)     \\
 \hline
 La$_{0.67}$Ca$_{0.33}$Mn$_{0.9}$Ga$_{0.1}$O$_{3}$\cite{PhysRevB.70.104417}MAP  &  0.380(2) &  1.365(8)  \\
 La$_{0.67}$Ca$_{0.33}$Mn$_{0.9}$Ga$_{0.1}$O$_{3}$\cite{PhysRevB.70.104417}KF  &   0.387(6) &  1.362(2)  \\
 $q$ = 0.8 This work  &  0.382(3) &  1.387(7)     \\
 \hline
 La$_{0.67}$Ba$_{0.22}$Sr$_{0.11}$Mn$_{0.9}$Fe$_{0.1}$O$_{3}$\cite{BenJemaa}MAP  &   0.398(2) &  1.251(5)   \\
 La$_{0.67}$Ba$_{0.22}$Sr$_{0.11}$Mn$_{0.9}$Fe$_{0.1}$O$_{3}$\cite{BenJemaa}KF  &   0.395(3) &  1.247(3)   \\
 $q$ = 0.7 This work  &  0.395(3) &  1.424(7)     \\
 \hline
 La$_{0.67}$Ba$_{0.22}$Sr$_{0.11}$Mn$_{0.8}$Fe$_{0.2}$O$_{3}$\cite{BenJemaa}MAP  &   0.411(1) &  1.241(4)   \\
 La$_{0.67}$Ba$_{0.22}$Sr$_{0.11}$Mn$_{0.8}$Fe$_{0.2}$O$_{3}$\cite{BenJemaa}KF  &   0.394(3) &  1.292(3)   \\
 $q$ = 0.7 This work  &  0.395(3) &  1.424(7)     \\
 \hline
 Pr$_{0.7}$Pb$_{0.3}$MnO$_{3}$\cite{PhysRevB.75.024419}MAP  &   0.404(6) &  1.354(20)     \\
 Pr$_{0.7}$Pb$_{0.3}$MnO$_{3}$\cite{PhysRevB.75.024419}KF  &   0.404(1) &  1.357(6)     \\
 $q$ = 0.65 This work  &  0.402(3) &  1.446(7)     \\
 \hline
 La$_{0.75}$(Sr,Ca)$_{0.25}$Mn$_{0.9}$Ga$_{0.1}$O$_{3}$\cite{BenJemaa}MAP  &   0.420(5) &  1.221(2)   \\
 La$_{0.75}$(Sr,Ca)$_{0.25}$Mn$_{0.9}$Ga$_{0.1}$O$_{3}$\cite{BenJemaa}KF  &   0.428(5) &  1.286(4)   \\
 $q$ = 0.55 This work  &  0.421(3) &  1.503(7)     \\
 \hline
 Pr$_{0.5}$Sr$_{0.5}$MnO$_{3}$\cite{PhysRevB.79.214426}MAP  &   0.443(2) &  1.339(6)     \\
 Pr$_{0.5}$Sr$_{0.5}$MnO$_{3}$\cite{PhysRevB.79.214426}KF  &   0.448(9) &  1.334(10)     \\
 $q$ = 0.45 This work  &  0.450(3) &  1.585(8)     \\
 \hline
 \hline
\end{tabular}
\label{tableexponentsE3N3}
\end{table}

\begin{table}[H]
\caption{Results for both static nonextensive critical exponents to $3$d ($\epsilon = 1$) nonextensive materials ($q \neq 1$) Ising ($N = 1$) systems, obtained from experiment through Modified Arrott (MA) plots \cite{PhysRevLett.19.786}, Kouvel-Fisher (KF) method \cite{PhysRev.136.A1626} and NSFT of this work.}
\begin{tabular}{ p{5.75cm}p{1.3cm}p{1.2cm}  }
 \hline
 \hline
  & $\beta_{q}$ & $\gamma_{q}$    \\
 \hline
 La$_{0.8}$Sr$_{0.2}$MnO$_{3}$\cite{J.Appl.Phys.Vasiliu-Doloca}  &   0.290(10) &        \\
 $q$ = 1.5 This work  &  0.308(1) &  1.190(3)     \\
 \hline
 Nd$_{0.55}$Sr$_{0.45}$Mn$_{0.98}$Ga$_{0.02}$O$_{3}$\cite{YU2018393}  & 0.308(10) &  1.197 \\
 $q$ = 1.5 This work  &  0.308(1) &  1.190(3)     \\
 \hline
 Pr$_{0.6}$Sr$_{0.4}$MnO$_{3}$\cite{PhysRevB.92.024409}MAP  &   0.315(0) &  1.095(7)      \\
 Pr$_{0.6}$Sr$_{0.4}$MnO$_{3}$\cite{PhysRevB.92.024409}KF  &   0.312(2) &  1.106(5)      \\
 $q$ = 1.2 This work  &  0.316(1) &  1.215(3)     \\
 \hline
 La$_{0.8}$Ca$_{0.2}$MnO$_{3}$\cite{ZHANG2013146}KF  &   0.316(7) &  1.081(36)      \\
 $q$ = 1.2 This work  &  0.316(1) &  1.215(3)     \\
 \hline
 La$_{0.7}$Ca$_{0.3}$Mn$_{0.85}$Ni$_{0.15}$O$_{3}$\cite{PHAN201440}MAP  &   0.320(9) &  0.990(82)      \\
 $q$ = 1.1 This work  &  0.320(1) &  1.227(3)     \\
 \hline
 Nd$_{0.6}$Sr$_{0.4}$MnO$_{3}$\cite{RSCAdvJeddi}MAP  &   0.320(6) &  1.239(2)      \\
 Nd$_{0.6}$Sr$_{0.4}$MnO$_{3}$\cite{RSCAdvJeddi}KF  &   0.323(2) &  1.235(4)      \\
 $q$ = 1.1 This work  &  0.320(1) &  1.227(3)     \\
 \hline
 Nd$_{0.6}$Sr$_{0.4}$MnO$_{3}$\cite{PhysRevB.92.024409}MAP  &   0.321(3) &  1.183(17)      \\
 Nd$_{0.6}$Sr$_{0.4}$MnO$_{3}$\cite{PhysRevB.92.024409}KF  &   0.308(4) &  1.172(11)      \\
 $q$ = 1.1 This work  &  0.320(1) &  1.227(3)     \\
 \hline
 Nd$_{0.67}$Ba$_{0.33}$MnO$_{3}$\cite{HCINI20152042}MAP  &   0.325(4) &  1.248(19)      \\
 Nd$_{0.67}$Ba$_{0.33}$MnO$_{3}$\cite{HCINI20152042}KF  &   0.326(5) &  1.244(33)      \\
 $q$ = 1.05 This work  &  0.322(1) &  1.234(3)     \\
 \hline
 $q$ = 1.0 \cite{GHODHBANE2013558} &  0.325(2) &  1.241(2)     \\
 \hline
 La$_{0.65}$Bi$_{0.05}$Sr$_{0.3}$MnO$_{3}$\cite{Phys.SolidStateBaazaoui}MAP  &   0.335(3) &  1.207(20)      \\
 La$_{0.65}$Bi$_{0.05}$Sr$_{0.3}$MnO$_{3}$\cite{Phys.SolidStateBaazaoui}KF  &   0.316(7) &  1.164(20)      \\
 $q$ = 0.8 This work  &  0.339(1) &  1.279(3)     \\
 \hline
 La$_{0.65}$Bi$_{0.05}$Sr$_{0.3}$Mn$_{0.94}$Ga$_{0.06}$O$_{3}$\cite{Phys.SolidStateBaazaoui}MAP & 0.334(4) & 1.192(8) \\
 La$_{0.65}$Bi$_{0.05}$Sr$_{0.3}$Mn$_{0.94}$Ga$_{0.06}$O$_{3}$\cite{Phys.SolidStateBaazaoui}KF  & 0.307(8) & 1.138(5) \\
 $q$ = 0.8 This work  &  0.339(1) &  1.279(3)     \\
 \hline
 \hline
 \end{tabular}
\label{tableexponentsE3N1}
\end{table}
The physical interpretation for these results is the following: the critical indices values quantify how much some physical quantity diverges near the critical point. For example, consider a physical quantity like susceptibility, whose inverse  measures how much the material is susceptible to changes in the magnetic field. It diverges stronger (weaker) than in the extensive situation as the corresponding critical exponent $\gamma$ assumes higher (lower) values. So higher (lower) values of the critical indices indicate systems more (less) susceptible to magnetic field changes. Then higher (lower) values of $q$ (see Tables \ref{tableexponentsE3N3}-\ref{tableexponentsE3N1}) are accompanied by lower (higher) values of the critical indices and represent systems interacting stronger (weaker) than in the extensive case. The $q$-parameter has been related to the microscopic properties of specific systems, for example in grain displacements for a full quasistatic plane shear \cite{PhysRevLett.115.238301} and inhomogeneities \cite{SoaresPinto2007}. Now by generalizing for all effects occurring in manganites, as inhomogeneities \cite{SoaresPinto2007}, defects and impurities \cite{Magnetochemistry.Turki,KHELIFI2014149,PhysRevB.75.024419,OMRI20123122,Ghosh_2005,doi:10.1063/1.2795796,GHODHBANE2013558,J.Appl.Phys.A.Berger,PhysRevB.68.144408,BenJemaa,PhysRevB.70.104417,PhysRevB.79.214426,J.Appl.Phys.Vasiliu-Doloca,YU2018393,PhysRevB.92.024409,ZHANG2013146,PHAN201440,RSCAdvJeddi,HCINI20152042,Phys.SolidStateBaazaoui} etc. and the competition among them, this means that $q$ is interpreted as one encoding some effective stronger (weaker) interaction for $1 < q < 2$ ($0 < q < 1$) than in the extensive situation. This interaction is effective because it is a result of the various effects aforementioned due to crystal imperfections. All these effects, acting in a complex manner, lead to some stronger (weaker) effective interaction when $q$ increases (decreases), among the constituents of the system, resulting in higher (lower) values of $T_{c}$ (see Table \ref{tablecriticalTC}) when $q$ increases (decreases). For destroying the order of systems interacting stronger (weaker), it is needed to furnish more (less) energy than in the extensive case, then the critical temperature increases (decreases). Thus the present field-theoretic approach of this Letter is suitable for describing the critical behavior of nonconventional or complex systems and is the first one proposed in literature, for our knowledge. Another important point to analyze is: consider the $q$-entropy $S_{q}$ for two-independent subsystems \cite{Tsallis1988}, namely $S_{q}(AB) = S_{q}(A) + S_{q}(B) + (1 - q)S_{q}(A)S_{q}(B)/k_{B}$. When $q > 1$ ($q < 1$) we have $S_{q}(AB) < S_{q}(A) + S_{q}(B)$ ($S_{q}(AB) > S_{q}(A) + S_{q}(B)$) and the entropy decreases (increases), which characterizes subextensivity (superextensivity) \cite{Tsallis1988}. In fact, when $q > 1$ ($q < 1$) the system presents effective stronger (weaker) interactions and the energy of the system decreases (increases). These results are consistent with the fundamental physical law which asserts that the entropy of a given system is some increasing function of its energy \cite{Callen}. Another way of seeing that is to consider the leading nonextensive contribution, valid for $|1 - q| < 1$, to the behavior of the system (with the energy $E$ in units of $k_{B}T$) from $e_{q}^{-E} \approx e^{-E}\left[1 - \frac{1}{2}(1 - q)E^{2}\right] \approx e^{-[E + \frac{1}{2}(1 - q)E^{2}]}$. The effective energy $E + \frac{1}{2}(1 - q)E^{2}$ decreases (increases) for $q > 1$ ($q < 1$). So, once again, both effective entropy and energy, consistently, decrease (increase) for $q > 1$ ($q < 1$). Since the last argument is applied for some general entropy $S_{q}$ and energy $E$, it can be generalized for every physical complex system.

\begin{table}[t]
\caption{Results for the critical temperature to $3$d ($\epsilon = 1$) nonextensive materials ($q \neq 1$) Heisenberg ($N = 3$) systems, obtained from experiment through Modified Arrott (MA) plots \cite{PhysRevLett.19.786} and Kouvel-Fisher (KF) methods.}
\begin{tabular}{ p{5.7cm}p{1.5cm}p{1.0cm}  }
 \hline
 \hline
  & $\beta_{q}$ & $T_{c,\hspace{.5mm}q}(K)$    \\
 \hline
 La$_{0.67}$Sr$_{0.33}$MnO$_{3}$\cite{MNEFGUI2014193}MAP  &   0.333(8) &  350    \\ 
 $q$ = 1.9 This work  &  0.334(3) &       \\
 \hline
 La$_{0.67}$Sr$_{0.33}$Mn$_{0.95}$V$_{0.05}$O$_{3}$\cite{MNEFGUI2014193}MAP  &   0.358(5) &  326  \\
 $q$ = 1.1 This work  &  0.360(3) &       \\
 \hline
 La$_{0.67}$Sr$_{0.33}$Mn$_{0.95}$V$_{0.10}$O$_{3}$\cite{MNEFGUI2014193}MAP  &   0.367(3) &  301  \\
 $q$ = 0.95 This work  &  0.369(3) &       \\
 \hline
 La$_{0.67}$Sr$_{0.33}$Mn$_{0.95}$V$_{0.15}$O$_{3}$\cite{MNEFGUI2014193}MAP  &   0.375(3) &  290  \\
 $q$ = 0.9 This work  &  0.373(3) &       \\
 \hline
 \hline
 La$_{0.67}$Ba$_{0.22}$Sr$_{0.11}$MnO$_{3}$\cite{BenJemaa}MAP  &   0.378(3) &  343   \\
 La$_{0.67}$Ba$_{0.22}$Sr$_{0.11}$MnO$_{3}$\cite{BenJemaa}KF  &   0.386(6) &   342   \\
 $q$ = 0.85 This work  &  0.378(3) &       \\
 \hline
 La$_{0.67}$Ba$_{0.22}$Sr$_{0.11}$Mn$_{0.9}$Fe$_{0.1}$O$_{3}$\cite{BenJemaa}MAP  &   0.398(2) &  191   \\
 La$_{0.67}$Ba$_{0.22}$Sr$_{0.11}$Mn$_{0.9}$Fe$_{0.1}$O$_{3}$\cite{BenJemaa}KF   &   0.395(3) &  189   \\
 $q$ = 0.7 This work  &  0.395(3) &       \\
 \hline
 La$_{0.67}$Ba$_{0.22}$Sr$_{0.11}$Mn$_{0.8}$Fe$_{0.2}$O$_{3}$\cite{BenJemaa}MAP  &   0.411(1) &  130   \\
 La$_{0.67}$Ba$_{0.22}$Sr$_{0.11}$Mn$_{0.8}$Fe$_{0.2}$O$_{3}$\cite{BenJemaa}KF   &   0.394(3) &  139   \\ 
 $q$ = 0.7 This work  &  0.395(3) &       \\
 \hline
 \hline
 La$_{0.7}$Ba$_{0.3}$MnO$_{3}$\cite{KHELIFI2014149}MAP  &   0.341(3)  &  339     \\
 La$_{0.7}$Ba$_{0.3}$MnO$_{3}$\cite{KHELIFI2014149}KF  &   0.357(160) &  340      \\
 $q$ = 1.5 This work  &  0.343(3) &      \\
 \hline
 La$_{0.6}$Pr$_{0.1}$Ba$_{0.3}$MnO$_{3}$\cite{KHELIFI2014149}MAP  &   0.396(16)   &  321     \\
 La$_{0.6}$Pr$_{0.1}$Ba$_{0.3}$MnO$_{3}$\cite{KHELIFI2014149}KF   &   0.391(2)    &  321      \\
 $q$ = 0.7 This work  &  0.395(3) &       \\
 \hline
 La$_{0.5}$Pr$_{0.2}$Ba$_{0.3}$MnO$_{3}$\cite{KHELIFI2014149}MAP  &   0.494(10)   &  304     \\
 La$_{0.5}$Pr$_{0.2}$Ba$_{0.3}$MnO$_{3}$\cite{KHELIFI2014149}KF   &   0.491(23)   &  304      \\
 $q$ = 0.35 This work  &  0.496(4) &       \\
 \hline
 \hline
 \end{tabular}
\label{tablecriticalTC}
\end{table}

\par \textit{Some other nonextensive models}.---Once we have validated experimentally NSFT, now we present results for some other models:

\par For both nonextensive percolation  \cite{Bonfirm_1981} ($\alpha = -1$ and $\beta = -2$) and Lee-Yang edge singularity  \cite{Bonfirm_1981} ($\alpha = -1$ and $\beta = -1$) in dimensions $d = 6 - \epsilon$ we obtain, $\eta_{q} = \eta + (q - 1)\frac{8\alpha\beta}{3(\alpha - 4\beta)[\alpha - 4\beta(2q - 1)]}\epsilon$, $\nu_{q}^{-1} = \nu^{-1} + (q - 1)\frac{40\alpha\beta}{3(\alpha - 4\beta)[\alpha - 4\beta(2q - 1)]}\epsilon$, $\omega_{q} = \omega$.
In the case of Lee-Yang edge singularity, $\eta_{q}$ and $\nu_{q}$ are not independent \cite{Bonfirm_1981,PhysRevD.95.085001} but they are related by the relation relation $\nu_{q}^{-1} = (d - 2 + \eta_{q})/2$ \cite{Bonfirm_1981,PhysRevD.95.085001}. Once we have evaluated the nonextensive critical exponent $\eta_{q}$, by applying $\alpha = -1$ and $\beta = -1$, we can obtain the nonextensive critical index $\nu_{q}$ from the relation just mentioned as a function of $\eta_{q}$. The remaining nonextensive critical exponents can be computed through scaling relations among them \cite{PhysRevD.103.116024}.

\par The extensive $N$-component $\phi^{6}$ theory in $d = 3 - \epsilon$ \cite{STEPHEN197389,PhysRevE.60.2071} represents tricritical points for which the case $N = 2$ corresponds to $^{3}$He-$^{4}$He mixtures or antiferromagnets in the presence of a strong external field (metamagnets) \cite{Hager_2002}. The corresponding nonextensive critical exponents are given by $\eta_{q} = \eta + \frac{1 - q}{q}\frac{(N + 2)(N + 4)}{12(3N + 22)^{2}}\epsilon^{2}$, $\nu_{q} = \nu + \frac{1 - q}{q}\frac{(N + 2)(N + 4)}{3(3N + 22)^{2}}\epsilon^{2}$.
Actually, $\eta$ and $\nu$ are known up to six-loop order \cite{Hager_2002}.

\par Extensive long-range critical exponents for $N$-component Ising-like models whose interactions decay as $1/r_{ij}^{d + \sigma}$ are defined at three distinct regions \cite{BrezinEandParisiGandRicci-TersenghiF}. The corresponding nonextensive indices, in $d = 2\sigma - \varepsilon$, assume the form $\eta_{\sigma, \hspace{.5mm}q} = \eta_{q}, \eta_{\sigma}, 0$ and $\nu_{\sigma, \hspace{.5mm}q} = \nu_{q}, \nu_{\sigma}, \nu_{\sigma} + \frac{1 - q}{q}\frac{(N + 2)}{\sigma^{2}(N + 8)}\varepsilon$ for $\sigma > 2 - \eta_{q}$, $d/2 < \sigma < 2 - \eta_{q}$, $\sigma < d/2$, respectively.   
where $\eta_{\sigma} = 2 - \sigma$ for all loop orders \cite{LohmannMSladeGLallaceBC,SladeG} and $\nu_{\sigma}$ has been computed up to two-loop level \cite{PhysRevLett.29.917} in the interval $d/2 < \sigma < 2 - \eta_{q}$.


\par The extensive Gross-Neveu model in $d = 2 + \epsilon$ \cite{PhysRevD.10.3235} describes both $\psi$ and $\bar{\psi}$ Dirac massive fermions of mass $\mathcal{M}$. The corresponding nonextensive indices are given by $\eta_{\psi, \hspace{.5mm}q} = \eta_{\psi} + \frac{1 - q}{q}\frac{(2N - 1)}{8(N - 1)^{2}}\epsilon^{2}$, $\eta_{\mathcal{M}, \hspace{.5mm}q} = \eta_{\mathcal{M}} + \frac{1 - q}{q}\frac{\epsilon}{2(N - 1)}$, $\nu_{q} = \nu$.
The extensive indices have been computed up to four-loop order can be found in Ref. \cite{PhysRevD.94.125028}.

\par Extensive uniaxial systems with strong dipolar forces in the $z$-direction with Hamiltonian \cite{PhysRevB.13.251} $\mathcal{H} =  -\sum_{\vec{x}\vec{x}^{\prime}}\sum_{\mu\nu}S_{\vec{x}}^{\mu}S_{\vec{x}^{\prime}}^{\nu}\left(V_{\mu\nu}(\vec{x} - \vec{x}^{\prime}) +\frac{ \gamma\partial_{z}\partial_{z^{\prime}}}{|\vec{x} - \vec{x}^{\prime}|^{d - 2}}\right)$,
where $V_{\mu\nu}(\vec{x})$ is the short-range potential and $\gamma$ is a parameter for turning the intensity of the dipolar forces possess the nonextensive critical exponents in $d = 3 - \epsilon$ given by $\eta_{q} = \eta + \frac{1 - q}{q}\frac{4(N + 2)}{9(N + 8)^{2}}\epsilon^{2}$, $\nu_{q}^{-1} = \nu^{-1} - \frac{1 - q}{q}\frac{(N + 2)}{(N + 8)}\epsilon$.
For two-loop order, $\eta$ and $\nu$ are displayed in Ref. \cite{PhysRevB.13.251}.

\par The critical behavior of the nonextensive spherical model \cite{PhysRev.86.821} can be obtained by taking the limit $N\rightarrow\infty$ \cite{PhysRev.176.718} of the nonextensive O($N$)$_{q}$ model of this Letter. Taking such an limit, we obtain in $d = 4 - \epsilon$, $\eta_{q} = \eta$, $\nu_{q} = \nu + \frac{1 - q}{q}\frac{\epsilon}{4}$. 
In the present model, the extensive values are known in exact form. They are $\eta = 0$ and $\nu = 1/(2 - \epsilon)$ \cite{PhysRevLett.28.240}. Then $\eta_{q}$ and $\nu_{q}$ are exact within the approximation of the present work.

\par The extensive Lifshitz critical behavior \cite{PhysRevLett.35.1678,PhysRevB.67.104415,PhysRevB.72.224432,Albuquerque_2001,LEITE2004281,PhysRevB.61.14691,PhysRevB.68.052408,FARIAS,Borba,Santos_2014,deSena_2015,Santos_2019,Santos_20192,Leite_2022} is composed of the $m$-axial Lifshitz points \cite{PhysRevB.67.104415} an their generalized form for the higher character situation \cite{PhysRevB.72.224432}. For the latter, there are $d - m$-, $m_{2}$-,..., $m_{n}$-dimensional vectors, respectively. The nonextensive generic higher character Lifshitz anisotropic (computed in the orthogonal approximation) and isotropic (calculates in both orthogonal approximation and exactly) critical exponents can be written as $\eta_{n, \hspace{.5mm}q} = \eta_{n} + \frac{1 - q}{q}n\frac{(N + 2)}{2(N + 8)^{2}}\epsilon_{n}^{2}$, $\nu_{n, \hspace{.5mm}q} = \nu_{n} + \frac{1 - q}{q}\frac{(N + 2)}{4n(N + 8)}\epsilon_{n}$,
where $\epsilon_{L} = 4 + \sum_{n = 2}^{L}[(n - 1)/n]m_{n} - d$, $\eta_{n, \hspace{.5mm}q} = \eta_{n} + \frac{1 - q}{q}\frac{(N + 2)}{2n(N + 8)^{2}}\epsilon_{n}^{2}$, $\nu_{n, \hspace{.5mm}q} = \nu_{n} + \frac{1 - q}{q}\frac{(N + 2)}{4n^{2}(N + 8)}\epsilon_{n}$, 
where $\epsilon_{L} = 4n - d$, $\eta_{n, \hspace{.5mm}q} = \eta_{n} + \frac{1 - q}{q}\frac{(-1)^{n + 1}\Gamma(2n)^{2}(N + 2)}{\Gamma(n + 1)\Gamma(3n)(N + 8)^{2}}\epsilon_{n}^{2}$, $\nu_{n, \hspace{.5mm}q} = \nu_{n} + \frac{1 - q}{q}\frac{(N + 2)}{4n^{2}(N + 8)}\epsilon_{n}$, respectively.
In Ref. \cite{PhysRevB.72.224432}, the extensive critical exponents are obtained up to next-to-leading level.


\par The critical exponents of the nonextensive long-range $\lambda\phi^{3}$ theory \cite{PhysRevB.31.379} in $d = 3\sigma - \varepsilon$ can be expressed as $\eta_{\sigma , \hspace{.5mm}q} = \eta_{\sigma}$, $\nu_{\sigma, \hspace{.5mm}q}^{-1} = \nu_{\sigma}^{-1} - \frac{1 - q}{2q - 1}\frac{\alpha}{\beta}\varepsilon$.
The values of $\alpha$ and $\beta$ are $-1$, $-1$ and $-1$, $-2$ for the nonextensive Lee-Yang edge singularity problem and percolation \cite{Bonfirm_1981}, respectively. The extensive index value $\eta_{\sigma} = 2 - \sigma$ is exact \cite{PhysRevB.31.379}. So $\eta_{\sigma ,\hspace{.5mm}q}$ is exact within the approximation of the present work. The extensive indices were computed earlier up to two-loop order in Ref. \cite{PhysRevB.31.379}.

\par The extensive multicritical points of order $k$ \cite{C.ItzyksonJ.M.Drouffe} possess the critical dimension $d_{c} = \frac{2k}{k - 1}$. The corresponding nonextensive critical exponent is given by $\eta_{q} = \eta + \frac{1 - q}{q}4(k - 1)^{2}\frac{(k)!^{6}}{(2k!)^{3}}\epsilon^{2}$
The extensive index $\eta$ was computed in Ref. \cite{C.ItzyksonJ.M.Drouffe} up to $2k - 2$-loop order.

\par The extensive Gross-Neveu-Yukawa model \cite{ZINNJUSTIN1991105}, in $d = 4 - \epsilon$, describes the interaction of one scalar field $\phi$ and $N$ massless Dirac fermions $\psi$ and $\bar{\psi}$. The nonextensive critical exponents of the model can be written as \cite{ZINNJUSTIN1991105} $\eta_{\psi ,\hspace{.5mm}q} = \eta_{\psi} + (1 - q)\frac{2\epsilon}{(2N + 3)[2N + 1 + 2(2q - 1)]}$, $\eta_{\phi ,\hspace{.5mm}q} = \eta_{\phi} + (1 - q)\frac{8N\epsilon}{(2N + 3)[2N + 1 + 2(2q - 1)]}$, $\nu_{q}^{-1} = \nu^{-1} - \frac{A_{N,\hspace{.5mm}q}}{(2N + 3)[2N + 1 + 2(2q - 1)]}\epsilon$,   
where $A_{N,\hspace{.5mm}q} = (2N + 3)(R_{N,\hspace{.5mm}q}/6 + 2N) - [2N + 1 + 2(2q - 1)](R_{N}/6 + 2N)$ and $R_{N,\hspace{.5mm}q} = -[2(N - 1)(2q - 1) - 1] + \sqrt{[2(N - 1)(2q - 1) - 1]^{2} + 144N(2q - 1)(3q - 2)}$. In Ref. \cite{PhysRevD.96.096010}, $\eta_{\psi}$, $\eta_{\phi}$ and $\nu$ were computed up to four-loop order.

\par For short- and long-range directed percolation \cite{JANSSEN2005147,Tauber_2005} in $d = 4 - \epsilon$ and $d = 2\sigma - \varepsilon$, respectively, we have $\eta_{q} = \eta - \frac{1 - q}{2q -1}\frac{\epsilon}{3}$, $\eta_{\sigma ,\hspace{.5mm}q} = \eta_{\sigma} - \frac{1 - q}{2q -1}\frac{2\varepsilon}{7}$, $\nu_{q} = \nu + \frac{1 - q}{2q -1}\frac{\epsilon}{8}$, $\nu_{\sigma ,\hspace{.5mm}q} = \nu_{\sigma} + \frac{1 - q}{2q -1}\frac{4\varepsilon}{7\sigma^{2}}$, $z_{q} = z - \frac{1 - q}{2q -1}\frac{\epsilon}{6}$, $z_{\sigma ,\hspace{.5mm}q} = z_{\sigma} - \frac{1 - q}{2q -1}\frac{2\varepsilon}{7}$. 
The extensive indices were evaluated up to two- and one-loop order in Ref. \cite{JANSSEN2005147}, respectively.

\par The critical behavior of short- and long-range dynamic isotropic percolation \cite{JANSSEN2005147,Tauber_2005} at $d = 6 - \epsilon$ and $d = 3\sigma - \varepsilon$, respectively, is given by the following nonextensive critical exponents, namely $\eta_{q} = \eta - \frac{1 - q}{2q -1}\frac{2\epsilon}{21}$, $\eta_{\sigma ,\hspace{.5mm}q} = \eta_{\sigma} - \frac{1 - q}{2q -1}\frac{3\varepsilon}{4}$, $\nu_{q} = \nu + \frac{1 - q}{2q -1}\frac{5\epsilon}{42}$, $\nu_{\sigma ,\hspace{.5mm}q} = \nu_{\sigma} + \frac{1 - q}{2q -1}\frac{\varepsilon}{2\sigma^{2}}$, $z_{q} = z - \frac{1 - q}{2q -1}\frac{\epsilon}{3}$, $z_{\sigma ,\hspace{.5mm}q} = z_{\sigma} - \frac{1 - q}{2q -1}\frac{3\varepsilon}{8}$.
The extensive critical exponents were obtained up to two- and one-loop order in Ref. \cite{JANSSEN2005147}, respectively.


\par \textit{Conclusions}.---In summary, we have validated NSFT by comparing the critical exponents values obtained from experimental measurements for various manganites with those predicted by NSFT of this work displayed in Tables \ref{tableexponentsE3N3}-\ref{tableexponentsE3N1}. The agreement was satisfactory, once the margin of error was $< 5\%$ for the most of manganites employed and as better as $|1 - q| < 1$ which is the limit of validity of the theory. The $q$-parameter was interpreted as one encoding some effective stronger (weaker) interaction for $1 < q < 2$ ($0 < q < 1$) when compared with systems interacting conventionally. The effective character of this interaction is due to the fact that it emerges from various effects from crystal imperfections as inhomogeneities \cite{SoaresPinto2007}, defects and impurities \cite{Magnetochemistry.Turki,KHELIFI2014149,PhysRevB.75.024419,OMRI20123122,Ghosh_2005,doi:10.1063/1.2795796,GHODHBANE2013558,J.Appl.Phys.A.Berger,PhysRevB.68.144408,BenJemaa,PhysRevB.70.104417,PhysRevB.79.214426,J.Appl.Phys.Vasiliu-Doloca,YU2018393,PhysRevB.92.024409,ZHANG2013146,PHAN201440,RSCAdvJeddi,HCINI20152042,Phys.SolidStateBaazaoui} etc. and the competition among them. Such an interpretation is consistent with the fundamental physical law of thermodynamics which asserts that entropy is an increasing function of the energy \cite{Callen}. Although we have interpreted $q$ for the specific situation of continuous phase transitions in manganites, we can generalize this result for every physical complex system. The theory presented in this Letter is the first field-theoretic one that proposes to describe imperfect or complex systems, for our knowledge. It generalizes its extensive counterpart proposed by Kenneth Wilson \cite{PhysRevLett.28.240}, which can be obtained in the specific limit $q \rightarrow 1$. We hope that the field-theoretic approach presented in this Letter can shed light on future theoretical, computational and experimental studies on complex systems.


\par \textit{Acknowledgments}---PRSC would like to thank the Brazilian funding agencies CAPES and CNPq (Grant: Produtividade 307982/2019-0) for financial support. 

\bibliography{apstemplate}

\end{document}